\def\ga{\mathrel{\mathpalette\fun >}}
\def\fun#1#2{\lower3.6pt\vbox{\baselineskip0pt\lineskip.9pt
  \ialign{$\mathsurround=0pt#1\hfil##\hfil$\crcr#2\crcr\sim\crcr}}}
\relax
\documentstyle[12pt]{article}
\advance\textheight by \topskip
\begin{document}
\begin{flushright}
{SU-ITP-93-9}\\
gr-qc/9304015 \\

April 12, 1993
\end{flushright}
\vskip 1.5cm
\begin{center}
{\Large\bf STATIONARY~ UNIVERSE}\vskip 1.0cm
{\bf Andrei Linde}\footnote{On leave  from: Lebedev
Physical Institute, Moscow. \
 E-mail: linde@physics.stanford.edu}   {\bf  and Arthur
Mezhlumian}\footnote{On
leave from: Landau Institute
for Theoretical Physics, Moscow. \ E-mail:
arthur@physics.stanford.edu}
 \vskip 0.5cm
{\it Department of Physics, Stanford University, \\
\vskip 0.3cm
Stanford, CA 94305-4060}\\
\end{center}
\vskip 1.0cm


{\centerline{\large ABSTRACT}}
\begin{quotation}
\vskip -0.4cm
 If the Universe contains at least one inflationary domain with a
sufficiently
large and homogeneous scalar field $\phi$, then this domain
permanently
produces new inflationary domains of all possible types.  We show
that under
certain conditions this process of the self-reproduction of the
Universe  can
be described by a {\it stationary} distribution of probability, which
means
that  the fraction of the physical volume of the Universe in a state
with given
properties (with  given values of  fields,  with a given density of
matter,
etc.) does not depend on time.    This represents a strong deviation
of
inflationary cosmology from the standard Big Bang paradigm.

\end{quotation}
\newpage

 The first models of inflation were  based on the standard assumption
of the
Big Bang theory that the Universe was created at a single moment of
time in a
state with the Planck density, and that it was hot and large (much
larger than
the Planck scale $M_p^{-1}$) from the very beginning. The  success
of
inflation in solving internal problems of the Big Bang  theory
apparently
removed the last doubts concerning the  Big Bang cosmology.  It
remained almost
unnoticed that during the last ten years the inflationary theory  has
broken
the umbilical cord connecting it with the old Big Bang theory, and
acquired an
independent life of its own. For the practical purposes of
description of the
observable part of our Universe one may still speak about the Big
Bang.
However, if one tries to understand the beginning of the Universe, or
its end,
or its global structure, then some of the notions of the Big Bang
theory become
 inadequate.

For example, already in the first version of the   chaotic inflation
scenario
\cite{b17} there was no need to assume that the whole Universe
appeared from
nothing at a single moment of time associated with the Big Bang, that
the
Universe was hot from the very beginning and that the inflaton scalar
field
$\phi$ which drives inflation originally occupied the minimum of its
potential
energy. Later it was found that if the Universe contains at least one
inflationary domain of a size of horizon  (`$h$-region') with a
sufficiently
large and homogeneous scalar field $\phi$, then this domain will
permanently
produce new $h$-regions of a similar type. In other words,  instead
of a single
Big Bang producing a one-bubble Universe, we are speaking now about
inflationary bubbles producing new bubbles, producing new bubbles,
{\it ad
infinitum}. In this sense, inflation is not a short intermediate
stage of
duration $\sim 10^{-35}$ seconds, but a self-regenerating process,
which occurs
in some parts of the Universe even now, and which will continue
without end.
The most striking realization of this scenario  occurs in the context
of
chaotic inflation   \cite{b19}, but the basic features of this
scenario remain
valid in old inflation \cite{b51}, new inflation \cite{b52,b62} and
extended
inflation as well \cite{EtExInf}.

Thus, recent developments of inflationary theory have considerably
modified our
cosmological paradigm \cite{MyBook}.  Now  we must learn  how to
formulate
physical questions in the new context. For example, in a homogeneous
part of
the Universe there is a simple relation between  the density of
matter and
time. However, on a very  large scale the Universe becomes extremely
inhomogeneous. Its density, at the same `cosmic time', varies
anywhere from
zero to the Planck density. Therefore the question about the density
of the
Universe at the time $10^{10}$ years may not have any definite
answer. Instead
of  addressing such questions we should study the distribution of
probability
of finding a part of the Universe with  given properties, and   find
possible
correlations between these properties.

It is extremely complicated to describe an inhomogeneous Universe and
to find
the corresponding probability distribution.  Fortunately, there
exists a
particular
kind of stationarity of the process of the Universe self-reproduction
which
makes things  more regular. Due to the no-hair theorem for de Sitter
space, the
process of production of new inflationary domains occurs
independently of any
processes outside the horizon. This process depends only on the
values of the
fields inside each $h$-region   of  radius $H^{-1}$. Each time  a new
inflationary $h$-region is created during the Universe expansion, the
physical
processes inside this region  will depend only on the properties of
the fields
inside it, but  not  on the `cosmic time'  at which it was created.

In addition to this most profound stationarity,  there may also exist
some
simple stationary probability distributions which may allow us to
say, for
example,  what  the probability is of finding a given field $\phi$ at
a given
point.  To examine this possibility one should consider the
probability
distribution $P_c(\phi, t|\chi)$, which  describes the probability
of finding
the field $\phi$ at a given point at a time $t$, under the condition
that at
the time $t=0$ the field $\phi$ at this point was equal to $\chi$
\cite{b62,b60}. The same function may also describe the probability
that the
scalar field which at  time $t$ was equal to $\phi$, at some earlier
time $t=0$
was equal to $\chi$.

The probability distribution $P_c$ has been studied by many authors,
see e.g.
\cite{b19}, \cite{MyBook}--\cite{Bond}. Our  investigation of this
question has
shown that in all realistic inflationary models the probability
distribution
$P_c(\phi, t|\chi)$ is not stationary \cite{b19,MyBook}. The reason
is very
simple.   The  probability distribution $P_c$  is  in fact the
probability
distribution per  unit volume in {\it comoving coordinates} (hence
the index
$c$ in  $P_c$), which do  not change during expansion of the
Universe.  By
considering this probability distribution   we neglect the main
source of the
self-reproduction of inflationary domains, which is the exponential
growth of
their volume. Therefore, in addition to   $P_c$, we introduced the
probability
distribution $P_p(\phi, t|\chi)$, which describes the probability to
find a
given field configuration in a unit physical volume  \cite{b19}. In
the present
paper (see also \cite{LLMPascos} for a more detailed presentation) we
will show
that under certain conditions the stationary probability distribution
$P_p(\phi, t|\chi)$ does exist, and a typical relaxation time during
which the
distribution $P_p(\phi, t|\chi)$ approaches the stationary regime is
extremely
small.

First of all we should remember some details of stochastic approach
to
inflation.
Let us consider the simplest model of chaotic inflation based on the
theory of
a  scalar field $\phi$ minimally coupled to gravity, with the
effective
potential $V(\phi)$. If the classical field $\phi$ is sufficiently
homogeneous
in some domain of the Universe, then its behavior inside this domain
is
governed by the equations
\begin{equation}\label{E02}
\ddot\phi + 3H\dot\phi = -dV/d\phi\ ,
\end{equation}
\begin{equation}\label{E03}
H^2 + \frac{k}{a^2} = \frac{8\pi}{3M_p}\, \left(\frac{1}{2}
\dot\phi^2 + V(\phi)\right) \ .
\end{equation}
Here $H={\dot a}/a,\,  a(t)$ is the scale factor of the
 Universe, $k=+1, -1,$ or $0$ for a closed, open or flat Universe,
respectively. $M_p$ is the Planck mass, which we will put equal to
one in the
rest of the paper.

Investigation of these equations has shown that for many potentials
$V(\phi)$
(e.g., in all power-law $V(\phi)\sim \phi^n$ and exponential
$V(\phi)\sim
e^{\alpha \phi}$ potentials) there exists an intermediate asymptotic
regime of
slow rolling of the field $\phi$ and quasi-exponential expansion
(inflation)
of the Universe \cite{MyBook}. At this stage the Hubble parameter is
$H(\phi)=
\sqrt{8\pi V(\phi)/3}$.  In the theories  $V(\phi)\sim \phi^n$
inflation ends
at $\phi_e \sim n/12$. In the theory with $V(\phi)\sim e^{\alpha
\phi}$
inflation ends only if we bend the potential at some point $\phi_e$;
for
definiteness we will take $\phi_e = 0$ in this theory.

Inflation stretches all initial inhomogeneities. Therefore, if the
evolution of the Universe were governed solely by classical equations
of
motion, we would end up with an extremely smooth Universe with no
primordial
fluctuations to initiate the growth of galaxies.
Fortunately, new density perturbations are generated during inflation
due to
quantum effects. The wavelengths of all  vacuum
fluctuations of the scalar field $\phi$ grow exponentially in the
expanding
Universe. When the wavelength of any particular fluctuation becomes
greater
than $H^{-1}$, this fluctuation stops oscillating, and its amplitude
freezes at
some nonzero value $\delta\phi (x)$ because of the large friction
term
$3H\dot{\phi}$ in the equation of motion of the field $\phi$\@. The
amplitude
of this fluctuation then remains
almost unchanged for a very long time, whereas its wavelength grows
exponentially. Therefore, the appearance of such a frozen fluctuation
is
equivalent to the appearance of a classical field $\delta\phi (x)$
that does
not vanish after averaging over macroscopic intervals of space and
time.

Because the vacuum contains fluctuations of all wavelengths,
inflation leads to
the creation of more and more perturbations of the classical field
with
wavelengths greater than $H^{-1}$\@. The average amplitude of such
perturbations generated during a time interval $H^{-1}$ (in which the
Universe
expands by a factor of e) is given by
\begin{equation}\label{E23}
|\delta\phi(x)| \approx \frac{H}{2\pi}\ .
\end{equation}
The phases of each wave are random. Therefore,  the sum of all waves
at a given
point fluctuates and experiences Brownian jumps in all directions in
the space
of fields.

The standard way to describe  the stochastic behavior of the inflaton
field during the slow-rolling stage is to coarse-grain it over
$h$-regions
and consider the effective equation of motion of the long-wavelength
field \cite{b60}:
 \begin{equation} \label{m1}
\frac{d \phi}{dt} = -\, \frac{V'(\phi)}{3H(\phi)} +
\frac{H^{3/2}(\phi)}{2\pi}
\, \xi(t) \ .
\end{equation}
Here  $H= \sqrt{8\pi V/3}$\,, \, $\xi(t)$ is the effective white
noise
generated by quantum fluctuations,
which leads to the Brownian motion of the classical field $\phi$\@.

This Langevin equation leads to two stochastic equations for the
probability
distribution $P_c(\phi,t|\chi)$. The first one is called the backward
Kolmogorov equation,
\begin{equation} \label{b60b}
\frac{\partial  P_c(\phi,t|\chi)}{\partial t} =
 \frac{1}{2} \frac{H^{3/2}(\chi)}{2\pi}\, \frac{\partial
}{\partial\chi}
 \left( \frac{H^{3/2}(\chi)}{2\pi} \, \frac{\partial }{\partial\chi}
P_c(\phi,t|\chi)
\right)
 -  \frac{V'(\chi)}{3H(\chi)} \frac{\partial }{\partial\chi}
P_c(\phi,t|\chi)   \ .
\end{equation}
In this equation one considers the value of the field $\phi$ at the
time \, $t$
as a constant,  and finds the time dependence of the probability that
this
value  was reached  during the time \,  $t$ as a result of
diffusion of the
scalar field from different possible initial values $\chi \equiv
\phi(0)$.

The second equation is the adjoint to the first one; it is called the
forward
Kolmogorov equation, or the Fokker-Planck equation,
\begin{equation}\label{E3711}
\frac{\partial P_c(\phi,t|\chi)}{\partial t} =
 \frac{1}{2}  \frac{\partial }{\partial\phi}
 \left( \frac{H^{3/2}(\phi)}{2\pi}\, \frac{\partial }{\partial\phi}
\Bigl(
 \frac{H^{3/2}(\phi)}{2\pi}  P_c(\phi,t|\chi) \Bigr) +
\frac{V'(\phi)}{3H(\phi)} \, P_c(\phi,t|\chi)\right) \ ,
\end{equation}
This equation was derived in \cite{b60}, see also \cite{b62}.

One may try to find a stationary solution of equations (\ref{b60b}),
(\ref{E3711}), assuming that  $\frac{\partial
P_c(\phi,t|\chi)}{\partial t} =
0$.
The simplest stationary solution  (subexponential factors being
omitted) would
be
\begin{equation}\label{E38a} P_c(\phi,t|\chi) \sim
\exp\left({3\over 8 V(\phi)}\right)\cdot \exp\left(-{3\over 8
V(\chi)}\right) \
{}.
\end{equation}
This function is extremely interesting. Indeed, the first term  in
(\ref{E38a})
is equal to the square of the Hartle-Hawking wave function of the
Universe
\cite{HH}, whereas the second one gives the square of the tunneling
wave
function  \cite{tunnel}\hskip .1cm !

At  first glance, this result gives a direct confirmation and a
simple physical
interpretation of  both the Hartle-Hawking  wave function of the
Universe {\it
and} the tunneling wave function. However, in all realistic
cosmological
theories, in which $V(\phi)=0$ at its minimum, the Hartle-Hawking
distribution
$\exp\left({3\over 8 V(\phi)}\right)$ is not normalizable. The source
of this
difficulty can be easily
understood: any stationary distribution may exist only due
to  compensation of the classical flow of the field $\phi$
downwards to the minimum of $V(\phi)$ by the diffusion motion
upwards. However, diffusion of the field $\phi$ discussed
above exists only during inflation. Thus, there is no diffusion
motion upwards
from the region $\phi < \phi_e$. Therefore all solutions of  equation
(\ref{E3711}) with the proper  boundary conditions at $\phi = \phi_e$
(i.e. at
the end of inflation)  are non-stationary (decaying)  ${\cite{b19}}$.

The situation with the probability distribution $P_p$ is much more
interesting
and complicated.  As  was shown in \cite{b19} its behavior  depends
strongly on
  initial conditions. If  the distribution $P_p$   was initially
concentrated
at $\phi < \phi^*$, where $\phi^*$ is some critical value of the
field,  then
it moves towards small $\phi$ in the same way as $P_c$, i.e. it
cannot become
stationary. On the other hand, if the initial value of the field
$\phi$ is
larger than $\phi^*$, the distribution  moves towards larger and
larger values
of the field $\phi^*$, until it reaches the field $\phi_p$, at which
the
effective potential of the field becomes of the order of Planck
density $M_p^4$
(we will assume $M_p=1$ hereafter), where the standard methods of
quantum field
theory in a curved classical space are no longer valid.

Some further steps towards the solution of this problem were made  by
Nambu
and Sasaki $ {\cite{Nambu}}$ and Miji\'c $ {\cite{Mijic}}$. Their
papers
contain many important results and insights. However, Miji\'c $
{\cite{Mijic}}$
 did not have a purpose to obtain a complete expression for the
stationary
distribution $P_p(\phi,t|\chi)$\@. The corresponding expressions were
obtained
for various types of potentials $V(\phi)$ in $ {\cite{Nambu}}$.
Unfortunately,
according to ${\cite{Nambu}}$, the stationary distribution
$P_p(\phi,t|\chi)$
is almost entirely concentrated at $\phi \gg \phi_p$, i.e. at
$V(\phi) \gg 1$,
where the methods used in $ {\cite{Nambu}}$ are inapplicable.

We will continue this investigation by writing the system of
stochastic
equations for $P_p$. These equations can be obtained from eqs.
(\ref{b60b}),
(\ref{E3711}) by adding the term $3HP_p$, which appears due to the
growth of
physical volume of the Universe by the factor $1 + 3H(\phi)\, dt$
during each
time interval $dt$ \cite{Nambu}--\cite{MezhMolch},
\cite{LLMPascos}\@:
\begin{equation} \label{b60bx}
\frac{\partial  P_p}{\partial t} =
 \frac{1}{2} \frac{H^{3/2}(\chi)}{2\pi}\, \frac{\partial
}{\partial\chi}
 \left( \frac{H^{3/2}(\chi)}{2\pi} \, \frac{\partial
P_p}{\partial\chi}
\right)
 -  \frac{V'(\chi)}{3H(\chi)} \frac{\partial  P_p}{\partial\chi}
  +3H(\chi)  P_p \ ,
\end{equation}
\begin{equation}\label{E372}
\frac{\partial P_p}{\partial t} =
 \frac{1}{2}  \frac{\partial }{\partial\phi}
 \left( \frac{H^{3/2}(\phi)}{2\pi}\, \frac{\partial }{\partial\phi}
\Bigl(
 \frac{H^{3/2}(\phi)}{2\pi}  P_p \Bigr) +
\frac{V'(\phi)}{3H(\phi)} \, P_p\right)  +  3H(\phi)  P_p\ .
\end{equation}
To find solutions of these equations one must specify boundary
conditions. The
boundary conditions at the end of inflation follow from the
conservation of the
probability flux  \cite{LLMPascos}:
\begin{equation}\label{TunnBound}
\frac{\partial }{\partial\phi}P_p({\phi_{e}}) = - {3\over 4}\
{V'\over V} \
P_p({\phi_{e}}) \  ,~~~~~\left. \frac{\partial }{\partial\chi}\left(
P_p\
\exp\Bigl({3\over 8V(\chi)}\Bigr)\right)\right|_{\chi =
\phi_{\phantom{}_{e}}}=
0 \ .
\end{equation}
The last boundary condition is especially interesting, since it
indicates that
at least in the vicinity of $\chi = \phi_{e}$ the probability
distribution with
respect to $\chi$ looks as a square of the tunneling wave function.
However, we
have found solutions of  (\ref{b60bx}), (\ref{E372}) to be rather
stable with
respect to  modification of these boundary conditions, whereas the
conditions
at the Planck boundary $\phi = \phi_p$ do play a very important role.

Investigation of these conditions poses many difficult problems.
First of all,
our diffusion equations are based on the semiclassical approach to
quantum
gravity, which breaks down at  super-Planckian densities. Secondly,
the shape
of the effective potential may be strongly modified by quantum
effects at $\phi
\sim \phi_p$. Finally, our standard interpretation of  the
probability
distribution $P_c$ and $P_p$ breaks down at $V(\phi) > 1$, since at
super-Planckian densities the notion of a classical scalar field in
classical
space-time does not make much sense.

However, these problems by themselves suggest a  possible answer.
Inflation
happens only in  theories with very flat effective potentials. At the
Planck
density nothing can protect the effective potential from becoming
steep.
Hence, one may expect that inflation ceases to exist at $\phi >
\phi_p$, which
leads to the boundary condition
\begin{equation}\label{PlanckBound}
 P_p(\phi_p,t|\chi)= P_p(\phi,t|\chi_p) = 0 \ ,
\end{equation}
where $V(\phi_p) \equiv V(\chi_p) = O(1)$.

There is also another, much more general  reason to expect that
inflation kills
itself as the potential energy density approaches the Planck density
$V \sim
1$. Indeed, the amplitude of fluctuations of the scalar field
generated during
the typical time $\delta t = H^{-1}$ is given by $H/2\pi$, and their
typical
wavelength at that time is $O(H^{-1})$. This means that the energy
density
associated with the gradients of these perturbations is of the order
of $H^4
\sim V^2$. Thus, in the domains with $V>1$ the gradient energy
density  $\sim
V^2$ becomes larger than the potential energy density $V$. This
violates one of
the basic assumptions necessary for inflation in domains with $V \ga
1$. One
may expect also that large gradients of energy on the scale
comparable to the
scale of the horizon $H^{-1}$ lead to creation of black holes  rather
than to
the permanent self-reproduction of  inflationary $h$-regions.

Of course, one may argue that all our considerations do not make
sense at
densities larger than the Planck density. When the energy  density in
any
$h$-region approaches the Planck density, it may no longer be
described in
terms of classical space-time and should be just thrown away from our
consideration. In particular, its volume should not be considered as
contributing to the total volume of the Universe. Thus, such domains
should be
neglected in our definition of $P_p$\@. In this case the
distribution $P_p$ for the field $\phi$ will stop moving towards
higher values
of $\phi$ and will approach a stationary regime when this
distribution will be
shifted towards $\phi \sim \phi_p$.

What we are saying is even stronger. Even if one makes an attempt to
consider
the domains with $V > 1$ as a part of classical space-time,  many
parts of
these domains drop out from the process of inflation. This means that
the total
volume of inflationary $h$-regions cannot grow as fast as $e^{3Ht}$.

We do not know which of these arguments, if any, will survive in the
future
theory of all fundamental interactions. However, all these arguments
point out
in the same direction: For a phenomenological description of
stochastic
processes in classical space-time one should impose boundary
conditions of the
type of (\ref{PlanckBound}) which do  not permit penetration of
inflation deep
into the realm of super-Planckian densities. As we will show in
\cite{LLMPascos}, the exact form of these boundary conditions is not
very
important; most of them allow
the same class of solutions as the boundary condition
(\ref{PlanckBound}).
Therefore in this paper we will use these boundary conditions,
assuming for
definiteness that they are imposed at $V(\phi_p) \equiv V(\chi_p)
=1$.

One may try to obtain solutions of equations (\ref{b60bx}),
(\ref{E372}) in the
form of the following series of biorthonormal system of
eigenfunctions of the
pair of adjoint linear operators (defined by the left hand sides of
the
equations below):
\begin{equation} \label{eq14}
P_p(\phi,t|\chi) =
\sum_{s=1}^{\infty} { e^{\lambda_s t}\, \psi_s(\chi)\, \pi_s(\phi) }
\ .
\end{equation}
Indeed, this gives us a solution of eq. (\ref{E372}) if
\begin{equation} \label{eq15}
 \frac{1}{2} \frac{H^{3/2}(\chi)}{2\pi} \frac{\partial
}{\partial\chi}
 \left( \frac{H^{3/2}(\chi)}{2\pi} \frac{\partial }{\partial\chi}
\psi_s(\chi) \right)
 - \frac{V'(\chi)}{3H(\chi)} \frac{\partial }{\partial\chi}
\psi_s(\chi)
 + 3H(\chi)  \cdot \psi_s(\chi) =
\lambda_s \, \psi_s(\chi)  \ .
\end{equation}
and\begin{equation} \label{eq17}
\frac{1}{2}  \frac{\partial }{\partial\phi}
 \left( \frac{H^{3/2}(\phi)}{2\pi} \frac{\partial }{\partial\phi}
\left(
 \frac{H^{3/2}(\phi)}{2\pi} \pi_j(\phi) \right) \right)
 + \frac{\partial }{\partial\phi} \left( \frac{V'(\phi)}{3H(\phi)} \,
\pi_j(\phi) \right)
 + 3H(\phi)  \cdot \pi_j(\phi) =
\lambda_j \, \pi_j(\phi)  \ .
\end{equation}
The orthonormality condition reads
\begin{equation} \label{eq20}
\int_{\phi_e}^{\phi_p} { \psi_s(\chi) \, \pi_j(\chi) \, d\chi }
= \delta_{sj} \ .
\end{equation}

In our case (with regular boundary conditions) one can easily show
that the
spectrum of $\lambda_j$ is discrete and bounded from above. Therefore
the
asymptotic solution for $P_p(\phi,t|\chi)$ (in the limit $t
\rightarrow \infty$) is given by
\begin{equation} \label{eq22}
P_p(\phi,t|\chi) = e^{\lambda_1 t}\, \psi_1(\chi) \,
\pi_1(\phi)\, \cdot \left(1 + O\left( e^{-\left(\lambda_1 - \lambda_2
\right) t} \right) \right) \ .
\end{equation}
Here $\psi_1(\chi)$ is the only positive eigenfunction of eq.
(\ref{eq15}),
$\lambda_1$ is the corresponding (real) eigenvalue, and $\pi_1(\phi)$
 is the eigenfunction
of the conjugate operator (\ref{eq17}) with the
same eigenvalue $\lambda_1$\@. Note, that $\lambda_1$ is
the highest eigenvalue, $\mbox{Re} \left( \lambda_1 - \lambda_2
\right) > 0 $\@. This is the reason why the asymptotic equation
(\ref{eq22}) is valid at large $t$\@. We have found \cite{LLMPascos}
that
in realistic theories of inflation a typical time of
relaxing to the asymptotic
regime, $\Delta t \sim (\lambda_1 - \lambda_2)^{-1}$, is extremely
small. Typically it
is only about a few thousands Planck times, i.e. about $10^{-40}$
sec.
 This means that the normalized distribution
\begin{equation} \label{eq22aa}
\tilde{P}_p(\phi,t|\chi) = e^{-\lambda_1 t}
\,P_p(\phi,t|\chi)
\end{equation}
rapidly converges to the time-independent normalized distribution
\begin{equation} \label{eq22a}
\tilde{P}_p(\phi|\chi) \equiv
 \tilde{P}_p(\phi,t \rightarrow \infty|\chi) =  \psi_1(\chi) \,
\pi_1(\phi) \
{}.
\end{equation}
It is this stationary distribution that we were looking for.  Because
the
growing factor  $e^{-\lambda_1 t}$ is the same for all $\phi$ (and
$\chi$), one
can use $\tilde{P}_p$ instead of ${P}_p$ for calculation of all {\it
relative}
probabilities. In particular, $\tilde{P}_p(\phi|\chi)$ gives us the
fraction of
the volume of the Universe occupied by the  field $\phi$, under the
condition
that the corresponding part of the Universe at some time in the past
contained
the field $\chi$.  The
remaining problem is to find the functions $\psi_1(\chi)$ and
$\pi_1(\phi)$,
and to check that all assumptions about the boundary conditions which
we made on the way to eq. (\ref{eq22}) are actually satisfied.

We have solved this problem for chaotic inflation in a wide class of
theories
including the theories with polynomial and exponential effective
potentials
$V(\phi)$ and found the corresponding stationary distributions
\cite{LLMPascos}. Here we will present some of our results for the
theories
${\lambda\over 4}\phi^4$ and $V_o\,e^{\alpha\phi}$.

 Solution of equations (\ref{eq15}) and (\ref{eq17}) for
$\psi_1(\chi)$ and
$\pi_1(\phi)$ in the theory ${\lambda\over 4}\phi^4$  shows that
these
functions are extremely small at  $\phi\sim \phi_e$ and $\chi\sim
\chi_e$. They
grow at large $\phi$ and $\chi$,  then rapidly decrease, and vanish
at $\phi =
\chi=  \phi_p$. With a decrease of $\lambda$ the solutions become
more and more
sharply peaked near the Planck boundary. (The functions $\psi_1$ and
$\pi_1$
for the exponential potential have a similar behavior, but they are
less
sharply peaked near $\phi_p$.) A detailed discussion  of these
solutions will
be contained in \cite{LLMPascos}. The eigenvalues $\lambda_1$
corresponding to
different coupling constants $\lambda$ are given by the following
table:
\vspace{-0.5cm}
\begin{center}
\begin{tabular}{|c|c|c|c|c|c|c|c|}
\hline \hline
 $\lambda$ & $1$ & $10^{-1}$ & $10^{-2}$ & $10^{-3}$ & $10^{-4}$ &
$10^{-5}$ &
$10^{-6}$ \\
\hline
 $\lambda_1$ & 2.813 & 4.418 & 5.543 & 6.405 & 7.057 & 7.538 & 7.885
\\
\hline \hline
\end{tabular}
\end{center}
\vspace{-0.5cm}
One can find also the second eigenvalue $\lambda_2$. For example, for
$\lambda = 10^{-4}$ one gets  $\lambda_2=6.789$. This means that for
$\lambda =
10^{-4}$ the time of relaxation  to the stationary distribution is
$\Delta t
\sim (\lambda_1-\lambda_2)^{-1} \sim 4 M_p^{-1} \sim 10^{-42}$
seconds --- a
very short time indeed.

Note that the parameter $\lambda_1$ shows the speed of exponential
expansion of
the volume filled by a given field $\phi$. {\it This speed does not
depend on
the field $\phi$}, and has the same order of magnitude as the speed
of
expansion at the Planck density. Indeed, $\lambda_1$ should be
compared to
$3H(\phi)= 2\sqrt{6\pi V(\phi)}$, which is  equal to $2\sqrt{6\pi}$
at the
Planck density. It can be shown \cite{LLMPascos} that  in the limit
$\lambda
\to 0$ the eigenvalue $\lambda_1$ also becomes equal to
$2\sqrt{6\pi} \approx
8.681$. The meaning of this result is very simple: in the limit
$\lambda \to 0$
our solution becomes completely concentrated near the Planck
boundary, and
$\lambda_1$ becomes equal to $3H(\phi_p)$.

At  first glance, independence of the speed of expansion of volume
$e^{\lambda_1 t}$ on the value of the field $\phi$ may seem
counterintuitive.
The meaning of this result is that the domain filled with the field
$\phi$
gives the largest contribution to the growing volume of the Universe
if it
first diffuses towards the Planckian densities, spends there as long
time as
possible {\it expanding with  nearly Planckian speed}, and then
diffuses back
to its original value $\phi$.

But what  about the field $\phi$ which is already at the Planck
boundary? Why
do the corresponding domains not  grow exactly with the Planckian
Hubble
constant $H(\phi_p) = 2\sqrt{6\pi}/3$\,? It happens partially due to
diffusion
and slow rolling of the field towards smaller $\phi$. However, the
leading
effect  is the destructive diffusion towards the space-time foam with
$\phi >
\phi_p$. One may  visualize this process  by painting white all
domains with
$V(\phi) < 1$, and by painting black domains filled by space-time
foam with
$V(\phi) > 1$.  Then each time $H^{-1}(\phi_p)$ the volume of white
domains
with $\phi \sim \phi_p$ grows approximately $e^3$ times, but some
`black holes'
appear in these domains, and, as a result, the total volume of white
domains
increases only $e^{3\lambda_1/2\sqrt{6\pi}}$ times. This suggests (by
analogy
with \cite{ArVil})   calling the factor
$d_f = 3\lambda_1/2\sqrt{6\pi}$
`the fractal dimension of classical space-time', or  `the fractal
dimension of
the inflationary Universe'. (Note that
$d_f < 3$ for $\lambda \not = 0$; for example, $d_f = 2.6$ for
$\lambda =
10^{-5}$.)
However, one should keep in mind that the fractal structure of the
inflationary
Universe in the chaotic inflation scenario in general is more
complicated than
in the new or old inflation and cannot be completely specified just
by one
fractal dimension \cite{LLMPascos}.

The distribution $\tilde{P}_p(\phi | \chi)  =  \psi_1(\chi) \,
\pi_1(\phi)$
which we have obtained  does not depend on time $t$. However, in
general
relativity one may use many different time parametrizations, and the
same
physics can be described
differently in different `times'. One of the most natural choices of
time in
the context of stochastic approach to inflation is the time $\tau  =
\ln{{a\left(x, t \right) \over a(x,0)}} =  \int{H(\phi(x,t),t)\ dt}$
\cite{b60,Bond}.  Here $a\left(x, t \right)$ is a local value of the
scale
factor in the inflationary Universe. By using this time variable, we
were able
to obtain not only numerical solutions to the stochastic equations,
but also
simple asymptotic expressions describing these solutions. For
example, for the
theory ${\lambda\over 4 } \phi^4$ both the eigenvalue $\lambda_1$ and
the `fractal dimension' $d_f$ (which in this case refers both to the
Planck
boundary at $\phi_p$ and to the end of inflation at $\phi_e$) are
given by $d_f
= \lambda_1 \sim 3-1.1\, \sqrt \lambda$, and the stationary
distribution is
\begin{eqnarray}
\tilde{P}_p(\phi,\tau|\chi) &\sim &  \exp\Bigl(-{3\over
8V(\chi)}\Bigr)\,
\Bigl({1\over V(\chi)+0.4} - {1\over1.4}\Bigr)\, \cdot \,   \phi
\,\exp\Bigl(-{\pi\, (3-\lambda_1)\phi^2}\Bigr) \nonumber \\
 &\sim & \exp\Bigl(-{3\over 2 \lambda\, \chi^4}\Bigr)\, \Bigl({4\over
\lambda
\chi^4+1.6} - {1\over1.4}\Bigr)\, \cdot \,   \phi \, \exp\Bigl(- 3.5
\sqrt\lambda\phi^2\Bigr)\ .
\end{eqnarray}
Note that the first factor coincides with the square of the tunneling
wave
function \cite{tunnel}! This expression is valid  in the whole
interval from
$\phi_e$ to  $\phi_p$ and it correctly describes asymptotic behavior
of
$\tilde{P}_p(\phi,\tau|\chi)$ both at  $\chi \sim \chi_e$ and at
$\chi \sim
\chi_p$.

A similar investigation can be carried out for the theory $V(\phi) =
V_o\
e^{\alpha\phi}$. The corresponding solution is
\begin{eqnarray}
\tilde{P}_p(\phi,\tau|\chi) &\sim &  \exp\Bigl(-{3\over
8V(\chi)}\Bigr)\,
\Bigl({1\over V(\chi)} - {1}\Bigr)\, \cdot \,  \Bigl({1\over V(\phi)}
-
{1}\Bigr)\, V^{-1/2}(\phi)\ .
\end{eqnarray}
This expression gives a  rather good approximation for
$\tilde{P}_p(\phi,\tau|\chi)$ for all $\phi$ and  $\chi$.

The main result of our work is that under certain conditions the
properties of
our   Universe  can be described by a time-independent probability
distribution, which we have found for theories with polynomial and
exponential
effective potentials. A lot of work still has to be done to verify
this
conclusion, see  \cite{LLMPascos}. However, once this result is taken
seriously, one should consider its interpretation and  rather unusual
implications.

When  making cosmological observations, we study our part of the
Universe and
find that in this part inflation ended  about $t_e \sim 10^{10}$
years ago. The
standard assumption of the first models of inflation was that the
total
duration of the inflationary stage was  $\Delta t \sim 10^{-35}$
seconds. Thus
one could come to an obvious conclusion that our part of the Universe
was
created in the Big Bang, at the time  $t_e + \Delta t \sim 10^{10}$
years ago.
However, in our scenario the answer is quite different.

Let us consider an inflationary domain which gave rise to the process
of
self-reproduction of new inflationary domains. For illustrative
purposes, one
can visualize self-reproduction of inflationary domains  as a
branching
process, which gives a qualitatively correct description of  the
actual
physical process we consider. During this process, the first
inflationary
domain of  initial radius $\sim H^{-1}(\phi)$ within the time
$H^{-1}(\phi)$
splits into $e^{3} \sim 20$ independent inflationary domains of
similar size.
Each of them contains a slightly different field $\phi$,  modified
both by
classical motion down to the minimum of $V(\phi)$ and by
long-wavelength
quantum fluctuations of amplitude $\sim H/2\pi$. After the next time
step
$H^{-1}(\phi)$, which will be slightly different for each of these
domains,
they split again, etc. The whole process now looks like a branching
tree
growing from the first (root) domain. The radius of each branch is
given by
$H^{-1}$;  the total volume of all domains at any given time $t$
corresponds to
the `cross-section' of all branches of the tree at that time, and is
proportional to the number of branches. This volume  rapidly grows,
but when
calculating it, one should   take into account  that those branches,
in which
the field becomes  larger than $\phi_p$,  die and fall down from the
tree, and
each branch in which the field becomes  smaller than  $\phi_e$, ends
on an
apple (a part of the Universe where inflation ended and life became
possible).

One of our results is that even after we discard at each given moment
the dead
branches and the branches ended with apples, the total volume of live
(inflationary) domains will continue growing exponentially, as
$e^{\lambda_1
t}$. What is even more interesting, we have found that very soon the
portion of
branches with given properties (with given values of scalar fields,
etc.)
becomes time-independent. Thus, by observing any finite part of a
tree  at any
given time $t$ one cannot tell how old   the tree is.

To give a most dramatic representation of our conclusions, let us see
where
most of the apples  grow. This can be done simply by integrating
$e^{\lambda_1
t}$ from $t=0$ to $t = T$ and taking the limit as $T\to \infty$. The
result
obviously diverges at large $T$ as
 $\lambda_1^{-1}\, e^{\lambda_1 T}$, which
means that most  apples grow at an indefinitely large distance from
the root.
In other words, if we ask what is the total duration of inflation
which
produced a typical apple, the answer is that it is indefinitely
large.

This conclusion may seem very strange. Indeed, if one takes a typical
point in
the root domain, one can show that inflation at this point ends
within a finite
time $\Delta t \sim 10^{-35}$ seconds. This is a correct (though
model-dependent) result which can be confirmed by stochastic methods,
using the
distribution $P_c(\phi,\Delta t|\chi)$ \cite{b19}.  How could  it
happen that
the duration of inflation was any longer than $10^{-35}$ seconds?

The answer is related to the choice between $P_c$ and $P_p$, or
between roots
and fruits. Typical points in the root domain drop out from the
process of
inflation within $10^{-35}$ seconds. The number of those points which
drop out
from inflation at a much later stage is exponentially suppressed,
but they
produce the main part of the total volume of the Universe.
Note that the length of each particular branch continued back in time
may well
be finite \cite{Vil}. However, there is no upper limit to the length
of each
branch, and, as we have seen, the longest branches produce almost all
parts of
the Universe with properties similar to the properties of the part
where we
live now. Since by local observations we can tell nothing about our
distance in
time from the root domain, our probabilistic arguments suggest that
the root
domain is, perhaps, indefinitely far away from us. Moreover, nothing
in our
part of the Universe depends on the distance from the root domain,
and,
consequently, on the distance from the Big Bang.

Thus,  inflation  solves many problems of the Big Bang theory and
ensures that
this theory provides an excellent description of the local structure
of the
Universe. However,  after making all kinds of improvements of  this
theory, we
are now winding up with a model of a stationary Universe, in
which the
notion of the Big Bang    loses its dominant position, being removed
to the indefinite past.

The stochastic approach to inflation used in our work has an
intermediate
position between purely classical and purely quantum mechanical
approaches to
cosmology. In particular,   stationarity of our probability
distribution is
closely reminiscent of the time-independence of the wave function of
the
Universe in the Wheeler-DeWitt equation \cite{DeWitt}. We hope that
stochastic
methods  may show us a way towards a complete quantum mechanical
description of
the stationary  state of the Universe.

The authors are grateful to  D. Linde for the help with computer
calculations
and to  M.  Miji\' c, S. Molchanov, A. Starobinsky and
A. Vilenkin for valuable discussions. This research was supported in
part  by
the National Science Foundation grant PHY-8612280.


\begin{thebibliography}{999}
\bibitem{b17} A.D. Linde, Phys. Lett. {\bf 129B} (1983) 177.
\bibitem{b19} A.D. Linde, Phys. Lett. {\bf 175B} (1986) 395;~ A.S.
Goncharov,
A.D. Linde and V.F. Mukhanov, Int. J. Mod.
Phys. {\bf A2} (1987) 561.
\bibitem{b51} A.H. Guth, Phys. Rev. {\bf D23} (1981) 347;~ J.R.
Gott, Nature {\bf 295}
(1982) 304;~  K. Sato, H. Kodama, M. Sasaki and K. Maeda, Phys. Lett.
   {\bf B108} (1982) 35.
\bibitem{b52} P.J. Steinhardt, in: {\bf The Very Early Universe},
G.W.
Gibbons, S.W. Hawking, S. Siklos, eds., Cambridge U.P. Cambridge,
England (1982), p. 251;  A.D. Linde,  {\bf Nonsingular Regenerating
Inflationary Universe}, Cambridge University preprint (1982).
\bibitem{b62} A. Vilenkin, Phys. Rev. {\bf D27} (1983) 2848.
\bibitem{EtExInf} A.D. Linde, Phys. Lett.  {\bf B249} (1990) 18.
\bibitem{MyBook} A.D. Linde, {\bf Particle Physics and Inflationary
Cosmology} (Harwood, Chur, Switzerland, 1990).
\bibitem{b60} A.A. Starobinsky, in: {\bf Fundamental Interactions}
(MGPI Press, Moscow, 1984), p. 55;~ A.S. Goncharov and A.D. Linde,
Sov. J.
Part. Nucl. {\bf 17} (1986) 369;~A.A. Starobinsky, in: {\bf Current
Topics in
Field
Theory, Quantum Gravity and Strings}, Lecture Notes in Physics, eds.
H.J. de Vega and N. Sanchez (Springer, Heidelberg 1986) {\bf 206},
p. 107.
\bibitem{Bond} D.S. Salopek and J.R.
Bond,  Phys. Rev. {\bf D42} (1990) 3936;  {\it ibid} {\bf D43} (1991)
1005.
\bibitem{LLMPascos} A.D. Linde, D.A. Linde, A. Mezhlumian, {\bf From
the Big
Bang
Theory to the Theory of a Stationary Universe}, Stanford University
preprint,
to
appear;~ A.D. Linde and A.~Mezhlumian, in preparation.
\bibitem{HH} J.B. Hartle and S.W. Hawking, Phys. Rev. {\bf D28}
(1983) 2960.
\bibitem{tunnel} A.D. Linde, JETP {\bf 60}  (1984) 211; Lett. Nuovo
Cim.
{\bf 39}  (1984) 401; Ya.B. Zeldovich and A.A. Starobinsky, Sov.
Astron.
Lett. {\bf 10} (1984) 135; V.A. Rubakov, Phys. Lett. {\bf 148B}
(1984)  280;
A. Vilenkin, Phys. Rev. {\bf D30}  (1984)  549.
\bibitem{Nambu} Y. Nambu and M. Sasaki, Phys.Lett. {\bf B219} (1989)
240;~
Y.  Nambu, Prog. Theor. Phys. {\bf 81} (1989) 1037.
\bibitem{Mijic} M. Miji\' c, Phys. Rev. {\bf D42} (1990) 2469; Int.
J. Mod. Phys. {\bf A6} (1991)  2685.
\bibitem{ZelLin} Ya.B. Zeldovich and A.D. Linde, unpublished (1986).
\bibitem{MezhMolch} A. Mezhlumian and S.A. Molchanov,  Stanford
University preprint SU-ITP-92-32 (1992), available as
cond-mat/9211003.
\bibitem{ArVil} M. Aryal and A. Vilenkin, Phys. Lett. {\bf B199}
(1987) 351.
\bibitem{Vil} A. Vilenkin, Phys. Rev. {\bf D46} (1992) 2355.
\bibitem{DeWitt} J.A. Wheeler, in: {\bf Relativity, Groups and
Topology},
eds. C.M. DeWitt and J.A. Wheeler (Benjamin, New York, 1968); B.S.
DeWitt,
Phys. Rev. {\bf 160} (1967) 1113.
 \end{thebibliography}
\end{document}